# Channel Coding over Multiple Coherence Blocks with Queueing Constraints

Deli Qiao, Mustafa Cenk Gursoy, and Senem Velipasalar
Department of Electrical Engineering, University of Nebraska-Lincoln, Lincoln, NE 68588
Email: dqiao726@huskers.unl.edu, gursoy@engr.unl.edu, velipasa@engr.unl.edu

*Abstract*— This paper investigates the performance of wireless systems that employ finite-blocklength channel codes for transmission and operate under queueing constraints in the form of limitations on buffer overflow probabilities. A block fading model, in which fading stays constant in each coherence block and change independently between blocks, is considered. It is assumed that channel coding is performed over multiple coherence blocks. An approximate lower bound on the transmission rate is obtained from Feintein's Lemma. This lower bound is considered as the service rate and is incorporated into the effective capacity formulation, which characterizes the maximum constant arrival rate that can be supported under statistical queuing constraints. Performances of variable-rate and fixed-rate transmissions are studied. The optimum error probability for variable rate transmission and the optimum coding rate for fixed rate transmission are shown to be unique. Moreover, the tradeoff between the throughput and the number of blocks over which channel coding is performed is identified.

## I. INTRODUCTION

An important class of wireless systems (e.g., systems that support streaming or interactive video applications) operate under quality of service (QoS) constraints in the form of limitations on queueing delays or buffer overflows. A measure of the throughput under such constraints is effective capacity [1], [2]. Effective capacity of wireless systems has recently been studied in various setting (see e.g., [1]-[4] and references therein). In most prior work, the service rates supported by the wireless channel are assumed to be equal to the instantaneous channel capacity values and no decoding errors are considered. On the other hand, in practice, since finite blocklength codewords are employed, communication is performed at rates less than the channel capacity with nonzero probability of error. Recently, [5] has analyzed the performance of finite block-length codewords in the presence of statistical QoS constraints. However, in [5], coding is assumed to be performed over one coherence block in which the fading stays constant. In this paper, we consider a more general setting and assume that codewords are sent over multiple coherence blocks. Hence, each codeword experiences multiple fading realizations. Coding over multiple blocks generally improves the performance since codeword length can be increased and protection against severe fading can be provided as codewords see multiple channel states. However, coding over many blocks can also lead to long delays or buffer overflows. Therefore, it is of significant interest to analyze the throughput of channel coding over multiple coherence blocks in the presence of buffer constraints, and identify whether there exists an optimal number of blocks over which coding needs to be performed.

## II. SYSTEM MODEL

In this paper, we consider a block flat-fading channel, and assume that the fading coefficients stay constant for a coherence block of $n$ symbols and change independently from one block to another. The discrete-time input and output relationship in the $l^{th}$ block is given by

$$y_i = h_l x_i + w_i \quad i = 1, 2, \ldots, n \qquad (1)$$

where $x_i$ and $y_i$ are the complex-valued channel input and output, respectively, in the $i^{th}$ symbol duration, $h_l$ is the channel fading coefficient in the $l$th block, and $w_i$ is the circularly symmetric complex Gaussian noise with zero mean and variance $N_0$, i.e., $w_i \sim \mathcal{CN}(0, N_0)$. We assume that the receiver has perfect channel side information (CSI) and hence perfectly knows the realizations of the fading coefficients $\{h_l\}$. On the other hand, we consider both cases of perfect and no CSI at the transmitter.

The channel input is assumed to be subject to $\mathbb{E}\{|x_i|^2\} \le \mathcal{E}_s$. It is well-known that when the receiver has perfect CSI, the capacity achieving input for the above fading Gaussian channel is Gaussian distributed. Hence, we assume that $x_i \sim \mathcal{CN}(0, \mathcal{E}_s)$. Since the input and noise are Gaussian distributed, the output is also conditionally Gaussian, i.e., $y_i | h_l \sim \mathcal{CN}(0, \mathcal{E}_s |h_l|^2 + N_0)$. Moreover, $y_i | x_i, h_l \sim \mathcal{CN}(h_l x_i, N_0)$. We further assume that the input is independent and identically distributed (i.i.d.) i.e., $p_{x^n} = \prod_{i=1}^n p_{x_i}(x_i)$, which implies $p_{y^n | x^n, h_l} = \prod_{i=1}^n p_{y_i | x_i, h_l}(y_i | x_i, h_l)$, and $p_{y^n | h_l} = \prod_{i=1}^n p_{y_i | h_l}(y_i | h_l)$.

## III. PRELIMINARIES

### A. Effective Capacity

In [2], effective capacity is defined as the maximum constant arrival rate[1] that a given service process can support in order to guarantee a statistical QoS requirement specified by the QoS exponent $\theta$. If we define $Q$ as the stationary queue length, then $\theta$ is the decay rate of the tail distribution of the queue length $Q$:

$$\lim_{q \to \infty} \frac{\log P(Q \ge q)}{q} = -\theta. \qquad (2)$$

Therefore, for large $q_{\max}$, we have the following approximation for the buffer violation probability: $P(Q \ge q_{\max}) \approx e^{-\theta q_{\max}}$. Hence, while larger $\theta$ corresponds to more strict QoS constraints, smaller $\theta$ implies looser QoS guarantees.

---
[1]For time-varying arrival rates, effective capacity specifies the effective bandwidth of the arrival process that can be supported by the channel.

Similarly, if $D$ denotes the steady-state delay experienced in the buffer, then $P(D \geq d_{\max}) \approx e^{-\theta \delta d_{\max}}$ for large $d_{\max}$, where $\delta$ is determined by the arrival and service processes [3]. The effective capacity is given by

$$R_E(\theta) = -\lim_{t \to \infty} \frac{1}{\theta t} \log_e \mathbb{E}\{e^{-\theta S[t]}\} \quad \text{bits/s}, \quad (3)$$

where the expectation is with respect to $S[t] = \sum_{i=1}^{t} r_s[i]$, which is the time-accumulated service process. $\{r_s[i], i = 1, 2, \ldots\}$ denotes the discrete-time stationary and ergodic stochastic service process.

### B. Mutual Information Density and Channel Coding Rate

As detailed above, effective capacity is determined by specifying the service rate or equivalently the instantaneous transmission rate. We assume that the transmitter performs channel coding over $m$ coherence blocks where $m = 1, 2, \ldots$ Therefore, it sends codewords of length $nm$ and each codeword experiences $m$ independent channel conditions. An upper bound on the maximum decoding error probabilities of random codes of length $nm$ is given by Feinstein's Lemma [6], [7]:

$$\epsilon \leq P\left(\frac{1}{nm} i(x^{nm}; y^{nm}|h_1^m) \leq R + \gamma\right) + P(x^{nm} \notin S_{nm}) + e^{-nm\gamma} \quad (4)$$

where $\gamma > 0$ is an arbitrary constant, $S_{nm} = \left\{\frac{1}{nm} \sum_{i=1}^{nm} \mathbb{E}\{|x_i|^2\} \leq \mathcal{E}\right\}$ is the constraint set, $i(x^{nm}; y^{nm}|h_1^m)$ is the mutual information density conditioned on the fading coefficients $(h_1, h_2, \ldots, h_m)$ seen in $m$ coherence blocks. The conditional mutual information density is defined as

$$i(x^{nm}; y^{nm}|h_1^m) = \log_2 \frac{p(y^{nm}|x^{nm}, h_1^m)}{p(y^{nm}|h_1^m)}. \quad (5)$$

Next, we obtain an expression for the mutual information density of the considered channel and input models (i.e., fading Gaussian channel with Gaussian input), and derive, under certain assumptions, an approximate lower bound on the rates attained by coding over $m$ coherence blocks.

For the system model introduced in Section II, we have

$$\frac{1}{nm} i(x^{nm}; y^{nm}|h_1^m)$$

$$= \frac{1}{nm} \sum_{l=1}^{m} \sum_{i=(l-1)n+1}^{ln} i(x_i; y_i|h_l)$$

$$= \frac{1}{nm} \sum_{l=1}^{m} \sum_{i=(l-1)n+1}^{ln} \log_2 \frac{f_{y_i|x_i, h_l}(y_i|x_i, h_l)}{f_{y_i|h_l}(y_i, h_l)}$$

$$= \frac{1}{nm} \sum_{l=1}^{m} \sum_{i=(l-1)n+1}^{ln} \left(\log_2\left(1 + \frac{\mathcal{E}_s |h_l|^2}{N_0}\right) \right.$$
$$\left. + \frac{|y_i|^2 \log_2 e}{|h_l|^2 \mathcal{E}_s + N_0} - \frac{|y_i - h_l x_i|^2 \log_2 e}{N_0}\right)$$

$$= \frac{1}{m} \sum_{l=1}^{m} \log_2\left(1 + \frac{\mathcal{E}_s |h_l|^2}{N_0}\right)$$
$$+ \frac{\log_2 e}{nm} \sum_{l=1}^{m} \sum_{i=(l-1)n+1}^{ln} \left(\frac{|y_i|^2}{|h_l|^2 \mathcal{E}_s + N_0} - \frac{|y_i - h_l x_i|^2}{N_0}\right)$$

Denoting $\text{SNR} = \frac{\mathcal{E}_s}{N_0}$ and extending the results in [6] and [7], we can immediately show that $i(x^{nm}; y^{nm})/(nm)$ has the same distribution as the random variable [8]

$$\frac{1}{m} \sum_{l=1}^{m} \log_2(1 + \text{SNR}|h_l|^2) + \frac{\log_2 e}{nm} \sum_{l=1}^{m} \sqrt{\frac{\text{SNR}|h_l|^2}{1 + \text{SNR}|h_l|^2}} \sum_{i=1}^{n} w_{li}$$

where $w_{li}$'s are i.i.d. Laplace random variables, each with zero mean and variance 2. The sum of $nm$ i.i.d. Laplace random variables has a Bessel-K distribution [6] and generally is difficult to deal with directly. On the other hand, for large enough values of the blocklength $nm$, it can be well approximated by a Gaussian random variable [7]. Therefore, the mutual information density achieved with the codewords of length $nm$ spreading over $m$ coherence blocks can be approximated as

$$\frac{1}{nm} i(x^{nm}; y^{nm})$$
$$\sim \mathcal{CN}\left(\frac{1}{m} \sum_{l=1}^{m} \log_2(1 + \text{SNR} z_l), \frac{\log_2^2 e}{m} \sum_{l=1}^{m} \frac{2\text{SNR} z_l}{nm(1 + \text{SNR} z_l)}\right) \quad (6)$$

where we have defined $z_l = |h_l|^2$. With this approximation, the first probability expression on the right-hand side of (4) can be written in terms of the Gaussian $Q$-function:

$$P\left(\frac{1}{nm} i(x^{nm}; y^{nm}|h_1^m) \leq R + \gamma\right)$$
$$= Q\left(\frac{\frac{1}{m} \sum_{l=1}^{m} \log_2(1 + \text{SNR} z_l) - R - \gamma}{\sqrt{\frac{\log_2^2 e}{m} \sum_{l=1}^{m} \frac{2\text{SNR} z_l}{nm(1 + \text{SNR} z_l)}}}\right). \quad (7)$$

By noting that the $Q$-function is invertible, we can rewrite the upper bound in (4) as a lower bound on the instantaneous rate achieved by coding over $m$ coherence blocks:

$$R \geq \frac{1}{m} \sum_{l=1}^{m} \log_2(1 + \text{SNR} z_l) - \sqrt{\frac{\log_2^2 e}{m} \sum_{l=1}^{m} \frac{2\text{SNR} z_l}{nm(1 + \text{SNR} z_l)}}$$
$$\times Q^{-1}(\epsilon - P(x^{nm} \notin S_{nm}) - e^{-nm\gamma}) - \gamma \quad (8)$$

for any $\gamma > 0$. Although the above lower bound can also be used in the subsequent analysis, we opt to further simplify it to make the analysis more tractable analytically. For sufficiently large values of $nm$, the terms $P(x^{nm} \notin S_{nm})$ and $e^{-nm\gamma}$ become very small and can be neglected[2]. Moreover, since the lower bound holds for any $\gamma > 0$, we can see that an approximate lower bound for the transmission rate is

$$R \geq R_{l,\epsilon} = \frac{1}{m} \sum_{l=1}^{m} \log_2(1 + \text{SNR} z_l)$$
$$- \sqrt{\frac{\log_2^2 e}{m} \sum_{l=1}^{m} \frac{2\text{SNR} z_l}{nm(1 + \text{SNR} z_l)}} Q^{-1}(\epsilon) \quad (9)$$

---

[2]As $nm$ increases without bound, it can be easily seen that $P(x^{nm} \notin S_{nm})$ approaches zero by noting the fact that the codewords are generated according to $p_{x^{nm}} = \prod_{i=1}^{nm} p_{x_i}(x_i)$ where $p_{x_i}$ is the Gaussian distribution with zero mean and variance $\mathcal{E}$ and by applying the law of large numbers which tells us that the sample variance approaches the statistical variance $\mathcal{E}$.

where the notation $R_{l,\epsilon}$ is used to emphasize that this is a lower bound for rates achieved with decoding error probability $\epsilon$. Henceforth, the analysis is based on $R_{l,\epsilon}$.

## IV. EFFECTIVE THROUGHPUT WITH CHANNEL CODING OVER MULTIPLE COHERENCE BLOCKS

The rate lower bound in (9) gives a characterization of the tradeoffs and interactions between the instantaneous transmission rate, decoding error probability and the fading coefficients when channel coding is performed over multiple coherence blocks using finite blocklength codes. In particular, we note that $R_{l,e}$ is achieved with probability $1-\epsilon$. With probability $\epsilon$, decoding error occurs. We assume that the receiver reliably detect the errors, and apply a simple ARQ mechanism and sends a negative acknowledgement requesting the retransmission of the message in case of an erroneous reception. Therefore, the data rate is effectively zero when error occurs. Under this assumption, the service rate (in bits per $nm$ symbols) is

$$r_s = \begin{cases} 0, & \text{with probability } \epsilon \\ nmR_{l,\epsilon}, & \text{with probability } 1-\epsilon \end{cases} \quad (10)$$

Similarly as in [5], we obtain the following result on the effective rate by inserting the above service rate formulation into the definition in (3) and noting that the service rate varies independently for one sequence of $m$ blocks to another due to the block fading assumption. Since it characterizes the throughput achieved by transmitting at rates possibly below the channel capacity using finite blocklength codes, we refer to this throughput measure as effective rate rather than effective capacity in the remainder of the paper.

*Theorem 1:* The effective rate (in bits per channel use) at a given SNR, error probability $\epsilon$, codeword length $nm$, and QoS exponent $\theta$ is

$$\mathsf{R}_E(\theta) = -\frac{1}{\theta nm}\log_e \mathbb{E}_{\mathbf{z}}\left\{\epsilon + (1-\epsilon)e^{-\theta nmR_{l,\epsilon}}\right\} \quad (11)$$

where $R_{l,\epsilon}$ is given in (9), and $\mathbf{z} = (z_1, \ldots, z_m)$ is the vector composed of the channel states experienced in $m$ blocks.

The effective rate in (11) provides a lower bound on the throughput as a function of SNR, decoding error probability $\epsilon$, fading coefficients, coherence blocklength $n$, the number of blocks, $m$, over which coding is performed, and the QoS exponent $\theta$. The following result shows that given the other parameters, the effective rate is maximized at a unique decoding error probability. Note that using very strong codes and having small error probabilities in the transmission necessitates small transmission rates leading to small throughput. On the other hand, if higher transmission rates with relatively weak channel coding are preferred, then communication reliability degrades and more retransmissions are required again lowering the throughput.

*Theorem 2:* Given the values of $m > 0, n > 0, \theta > 0$ and SNR $> 0$, the function

$$\Psi(\epsilon) = \mathbb{E}\left\{\epsilon + (1-\epsilon)e^{-\theta nmR_{l,\epsilon}}\right\} \quad (12)$$

is strictly convex in $\epsilon$ and hence the optimal $\epsilon > 0$ that minimize $\Psi(\epsilon)$, or equivalently maximizes the effective throughput, is unique.

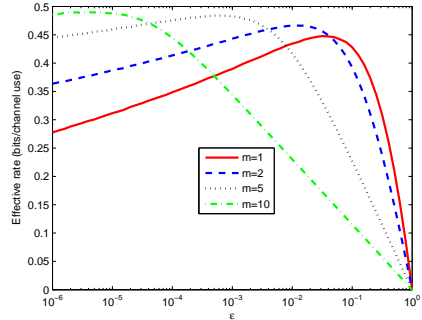

Fig. 1. The effective rate as a function of $\epsilon$. $n = 200$.

*Proof:* See Appendix A.

In Fig. 1, we plot the effective rate as a function of $\epsilon$ in the Rayleigh fading channel with $\mathbb{E}\{z\} = 1$. In the figure, we assume SNR $= 0$ dB, $\theta = 0.01$, and $n = 200$. We provided curves for different values of $m$. We can see that the effective throughput is indeed maximized at a unique $\epsilon$.

Another interesting tradeoff is the one between the throughput and the number of block over which coding takes place. Note that coding over multiple blocks is generally beneficial in terms of increasing the transmission rate on the average because transmitted codewords experience multiple channel fading realizations and may not get exceedingly affected by severe fading in one block. On the other hand, coding (with possible retransmissions in cases of decoding errors) over multiple blocks may lead to unacceptable delays in systems operating under buffer constraints captured by the QoS exponent $\theta$ in this paper. Therefore, coding over many blocks may lead to decreased throughput eventually. In Fig. 2, we plot the effective throughput as a function of $m$ for different $\theta$ values with fixed $\epsilon = 0.01$. We assume $n = 50$. In the figure, we observe that the optimal $m$ that maximizes the effective rate under a given $\epsilon$ varies with $\theta$. When $\theta = 0$ and therefore there are no buffer constraints, effective rate increases with increasing $m$. Coding over ever increasing number of blocks improves the performance. Indeed, as $m \to \infty$, effective rate approaches the ergodic capacity in the case of $\theta = 0$. However, we see a strikingly different behavior in the presence of QoS limitations. We note that for $\theta > 0$, effective rate is maximized at a finite value of $m$. Moreover, the optimal value of $m$ diminishes as $\theta$ increases. Therefore, coding over fewer blocks should be preferred under stringent buffer limitations.

In Fig. 3, we plot the optimal effective rate (optimized over the decoding error probability, $\epsilon$) as a function of $\theta$ for given $m$ values. We set $n = 50$. We find that for small $\theta$ values, having $m = 10$ achieves the highest effective rate, while as $\theta$ increases, having $m = 10$ starts providing the lowest effective throughput, due to similar reasons as outlined above.

### A. Fixed Rate Transmissions

Heretofore, we have implicitly assumed that the transmitter has perfect CSI and considered the scenario in which the transmitter employs variable-rate transmissions with rates characterized by $R_{l,\epsilon}$ given in (9). Note that in order to transmit at the rate $R_{l,\epsilon}$, the transmitter needs to know the fading coefficients $\{z_l\}_{l=1}^m$. A more practical scenario is the one in which the transmitter does not know the channel states and send the information at a fixed rate of $R$. Note that in this case, the decoding error probability varies with the fading coefficients in each set of $m$ blocks in contrast to being constant in the previous analysis. The codeword error probability for a given channel state $\mathbf{z}$ is

$$\epsilon(\mathbf{z}, R) = Q\left(\frac{\frac{1}{m}\sum_{l=1}^m \log_2(1+\text{SNR}z_l) - R}{\sqrt{\frac{1}{m}\sum_{l=1}^m \frac{2\text{SNR}z_l}{nm(1+\text{SNR}z_l)}\log_2 e}}\right) \quad (13)$$

obtained by using (9). The effective rate is then expressed as

$$\begin{aligned}
&\mathsf{R}_E(\theta, R) \\
&= -\frac{1}{\theta nm}\log_e \mathbb{E}_{\mathbf{z}}\left\{\epsilon(\mathbf{z}, R) + (1-\epsilon(\mathbf{z}, R))e^{-\theta nmR}\right\} \\
&= -\frac{1}{\theta nm}\log_e \mathbb{E}\left\{Q\left(\frac{\frac{1}{m}\sum_{l=1}^m \log_2(1+\text{SNR}z_l) - R}{\sqrt{\frac{1}{m}\sum_{l=1}^m \frac{2\text{SNR}z_l}{nm(1+\text{SNR}z_l)}\log_2 e}}\right)\right. \\
&\quad \left.+ \left(1 - Q\left(\frac{\frac{1}{m}\sum_{l=1}^m \log_2(1+\text{SNR}z_l) - R}{\sqrt{\frac{1}{m}\sum_{l=1}^m \frac{2\text{SNR}z_l}{nm(1+\text{SNR}z_l)}\log_2 e}}\right)\right)e^{-\theta nmR}\right\}.
\end{aligned}$$
(14)

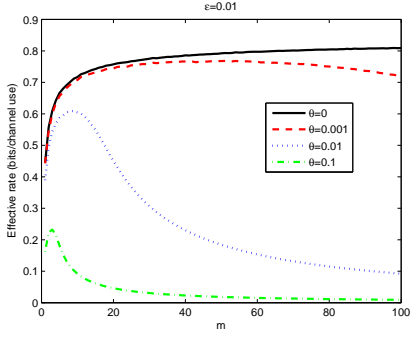

Fig. 2. The effective rate as a function of $m$. $n = 50$. $\epsilon = 0.01$.

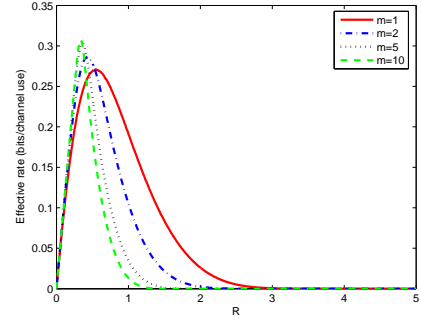

Fig. 4. The effective rate as a function of $R$. $n = 200$.

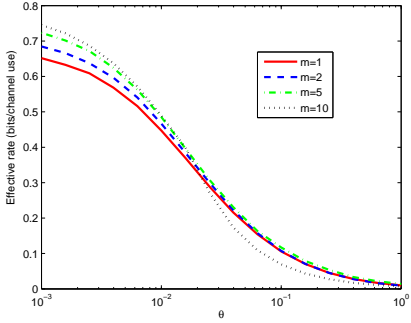

Fig. 3. The optimal effective rate vs. $\theta$. $n = 50$.

We have the following result.

*Theorem 3:* Assume that the values of $n$, $m$, $\theta > 0$ and SNR $> 0$ are fixed, then the function

$$\begin{aligned}
\Phi(R) = \mathbb{E}&\left\{Q\left(\frac{\frac{1}{m}\sum_{l=1}^m \log_2(1+\text{SNR}z_l) - R}{\sqrt{\frac{1}{m}\sum_{l=1}^m \frac{2\text{SNR}z_l}{nm(1+\text{SNR}z_l)}\log_2 e}}\right)\right. \\
&\left.+ \left(1 - Q\left(\frac{\frac{1}{m}\sum_{l=1}^m \log_2(1+\text{SNR}z_l) - R}{\sqrt{\frac{1}{m}\sum_{l=1}^m \frac{2\text{SNR}z_l}{nm(1+\text{SNR}z_l)}\log_2 e}}\right)\right)e^{-\theta nmR}\right\}
\end{aligned}$$

is minimized at a unique $R$ and hence the optimal $R$ that minimizes $\Phi(R)$ or equivalently maximizes the effective rate in (14) is unique.

*Proof:* See Appendix B.

In Fig. 4, we plot the effective rate as a function of the fixed transmission rate $R$ for different $m$ values. We assume that $\theta = 0.01$, $n = 200$. It is noted that there is a unique $R$ that maximizes the effective throughput. While we see in the figure that the optimal effective throughput increases as $m$ increases from 1 to 10, further numerical analysis showed results similar to those discussed in the variable-rate case, i.e., effective rate starts decreasing when $m$ is increased beyond a threshold value.

## V. CONCLUSION

We have analyzed the performance of channel coding over multiple blocks with possible decoding errors in the presence of queueing constraints. We have characterized the effective throughput. We have discussed two different transmission strategies. For the case when the transmission rate is varied and the error probability is kept fixed over different codewords, we have shown that the optimal error probability that maximizes the effective throughput is unique. Similarly, when the transmission rate is kept fixed, we have proved that the optimal rate that maximizes the effective throughput is also unique. Through numerical analysis, we have quantified the tradeoff between the throughput and the number of blocks over which coding is performed.

## APPENDIX

### A. Proof to Theorem 2

We first prove the following.

*Proposition 1:* The function

$$f(\epsilon) = (1-\epsilon)e^{-\theta nmR_{l,\epsilon}} \quad (15)$$

is strictly convex in $\epsilon$.

*Proof:* Denote
$$-\theta n m R_{l,\epsilon} = a Q^{-1}(\epsilon) + b \quad (16)$$

where we, from (9), define
$$a = \theta \sqrt{\sum_{l=1}^{m} \frac{2n\text{SNR}z_l}{1+\text{SNR}z_l} \log_2 e}, \quad b = -\theta n \sum_{l=1}^{m} \log_2(1+\text{SNR}z_l). \quad (17)$$

Note that $a > 0$ since $\text{SNR} > 0$, $z_l > 0$ and $\theta > 0$. Then we can rewrite (15) as
$$f(\epsilon) = (1-\epsilon)e^{aQ^{-1}(\epsilon)+b}. \quad (18)$$

The first and second derivative of $f(\epsilon)$ with respect to $\epsilon$ are
$$\dot{f}(\epsilon) = \left(a\dot{Q}^{-1}(\epsilon)(1-\epsilon) - 1\right)e^{aQ^{-1}(\epsilon)+b} \quad (19)$$
$$\ddot{f}(\epsilon) = \left(a(1-\epsilon)\left(\dot{Q}^{-1}(\epsilon)\right)^2 - 2\dot{Q}^{-1}(\epsilon) + (1-\epsilon)\ddot{Q}^{-1}(\epsilon)\right)$$
$$\times a e^{aQ^{-1}(\epsilon)+b} \quad (20)$$

where $\dot{Q}^{-1}(\epsilon)$ and $\ddot{Q}^{-1}(\epsilon)$ denote the first and second derivatives of $Q^{-1}(\epsilon)$ with respect to $\epsilon$. Note that for an invertible and differentiable function $g$, we have $g(g^{-1}(x)) = x$. Taking the derivative of both sides, we have
$$\dot{g}(g^{-1}(x))\dot{g}^{-1}(x) = 1 \Rightarrow \dot{g}^{-1}(x) = \frac{1}{\dot{g}(g^{-1}(x))} \quad (21)$$

where $\dot{g}(g^{-1}(x))$ denotes the first derivative of $g$ evaluated at $g^{-1}(x)$, and $\dot{g}^{-1}(x)$ is the derivative of $g^{-1}$ with respect to $x$. Noting that
$$Q(x) = \int_x^\infty \frac{1}{\sqrt{2\pi}} e^{-\frac{t^2}{2}} dt, \quad \dot{Q}(x) = -\frac{1}{\sqrt{2\pi}} e^{-\frac{x^2}{2}}, \quad (22)$$

we can derive the following
$$\dot{Q}^{-1}(\epsilon) = -\sqrt{2\pi} e^{-\frac{(Q^{-1}(\epsilon))^2}{2}}. \quad (23)$$

Note that $\dot{Q}^{-1}(\epsilon) < 0$ for $0 \leq \epsilon \leq 1$. Differentiating $\dot{Q}^{-1}(\epsilon)$ with respect to $\epsilon$, we have
$$\ddot{Q}^{-1}(\epsilon) = 2\pi Q^{-1}(\epsilon) e^{-Q^{-1}(\epsilon)^2}. \quad (24)$$

Next, we consider the following two cases:

1) $\epsilon < \frac{1}{2}$: We have $Q^{-1}(\epsilon) > 0$ for this case and hence $\ddot{Q}^{-1}(\epsilon) > 0$. Together with the fact that $\dot{Q}^{-1}(\epsilon) < 0$, we can immediately see that $\ddot{f}(\epsilon) > 0$ for $\epsilon < \frac{1}{2}$.

2) $\epsilon > \frac{1}{2}$: We have $Q^{-1}(\epsilon) < 0$ for this case. Substituting (23) and (24) to (20) and denoting $x = Q^{-1}(\epsilon)$, the expression inside the parentheses of (20) can be written as

$$a(1-\epsilon)\left(\dot{Q}^{-1}(\epsilon)\right)^2 - 2\dot{Q}^{-1}(\epsilon) + (1-\epsilon)\ddot{Q}^{-1}(\epsilon) \quad (25)$$
$$= a(1-\epsilon)2\pi e^{-(Q^{-1}(\epsilon))^2} + 2\sqrt{2\pi} e^{-\frac{(Q^{-1}(\epsilon))^2}{2}}$$
$$+ (1-\epsilon)2\pi Q^{-1}(\epsilon) e^{-Q^{-1}(\epsilon)^2} \quad (26)$$
$$= a(1-Q(x))2\pi e^{-\frac{x^2}{2}} + 2\sqrt{2\pi} e^{\frac{x^2}{2}} + (1-Q(x))2\pi x e^{-x^2} \quad (27)$$
$$= e^{\frac{x^2}{2}}\left(2\pi(1-Q(x))(x+a)e^{\frac{x^2}{2}} + 2\sqrt{2\pi}\right) \quad (28)$$
$$\geq e^{\frac{x^2}{2}}\left(2\pi(1-Q(x))xe^{\frac{x^2}{2}} + 2\sqrt{2\pi}\right) \quad (29)$$
$$\geq e^{\frac{x^2}{2}}\left(2\pi \frac{1}{\sqrt{2\pi}(-x)} e^{-\frac{x^2}{2}} x e^{\frac{x^2}{2}} + 2\sqrt{2\pi}\right) \quad (30)$$
$$= e^{\frac{x^2}{2}}(-\sqrt{2\pi} + 2\sqrt{2\pi}) = e^{\frac{x^2}{2}} \sqrt{2\pi} > 0 \quad (31)$$

where (29) follows from the facts that $a > 0$ and hence $x + a > x$. (30) is obtained from the following upper bound
$$1 - Q(x) = Q(-x) > \frac{1}{\sqrt{2\pi}(-x)} e^{-\frac{x^2}{2}} \text{ for } x < 0 \quad (32)$$

and the fact that $x < 0$ for this case, and hence multiplication of $x(1-Q(x))$ can be lowerbounded. Therefore, $\ddot{f}(\epsilon) > 0$ for $\epsilon > \frac{1}{2}$.

Also note that $\epsilon = \frac{1}{2}$ means $Q^{-1}(\epsilon) = 0$, so we have
$$a(1-\epsilon)\left(\dot{Q}^{-1}(\epsilon)\right)^2 - 2\dot{Q}^{-1}(\epsilon) + (1-\epsilon)\ddot{Q}^{-1}(\epsilon) \quad (33)$$
$$= a(1-\epsilon)2\pi e^{-(Q^{-1}(\epsilon))^2} + 2\sqrt{2\pi} e^{-\frac{(Q^{-1}(\epsilon))^2}{2}}$$
$$+ (1-\epsilon)2\pi Q^{-1}(\epsilon) e^{-Q^{-1}(\epsilon)^2} \quad (34)$$
$$= a\pi + 2\sqrt{2\pi} > 0 \quad (35)$$

and as a result $\ddot{f}(\epsilon) > 0$.

From the above discussion, we can find that $\ddot{f}(\epsilon) > 0$ for all $\epsilon \in [0,1]$. $f(\epsilon)$ is strictly convex in $\epsilon$. $\square$

Now, let $\psi(\epsilon) = \epsilon + (1-\epsilon)e^{-\theta n m R_{l,\epsilon}} = \epsilon + f(\epsilon)$. $\ddot{\psi}(\epsilon) = \ddot{f}(\epsilon) > 0$. Since the nonnegative weighted sum of strictly convex functions is a convex function [9], we can conclude that $\Psi(\epsilon)$ is convex.

## B. Proof to Theorem 3

First, for any given channel state pair $\mathbf{z} = (z_1, z_2, \ldots, z_m)$, we define
$$\mu = \frac{1}{m} \sum_{l=1}^{m} \log_2(1 + \text{SNR}z_l), \quad (36)$$
$$\delta = \sqrt{\frac{1}{m} \sum_{l=1}^{m} \frac{2\text{SNR}z_l}{nm(1+\text{SNR}z_l)} \log_2 e} \quad (37)$$

and note that $\mu > 0, \delta > 0$. We can find that $\Phi(0) = 1$, $\Phi(\infty) = 1$, and $\Phi(R) < 1$ for all $R \in (0, \infty)$. Note that
$$Q(x) = \int_x^\infty \frac{1}{\sqrt{2\pi}} e^{-\frac{t^2}{2}} dt. \quad (38)$$

The first and second derivatives of $\Phi(R)$ in $R$ are given by

$$\dot{\Phi}(R) = \mathbb{E}\left\{\frac{1}{\sqrt{2\pi}\delta}e^{-\frac{(\mu-R)^2}{2\delta^2}}\right\}\left(1-e^{-\theta nmR}\right)$$
$$-\theta nm\left(1-\mathbb{E}\left\{Q\left(\frac{\mu-R}{\delta}\right)\right\}\right)e^{-\theta nmR} \quad (39)$$

$$\ddot{\Phi}(R) = \mathbb{E}\left\{\frac{1}{\sqrt{2\pi}\delta}e^{-\frac{(\mu-R)^2}{2\delta^2}}\frac{\mu-R}{\delta^2}\right\}\left(1-e^{-\theta nmR}\right) + \theta nm e^{-\theta nmR}$$
$$\times\left(\mathbb{E}\left\{\frac{2}{\sqrt{2\pi}\delta}e^{-\frac{(\mu-R)^2}{2\delta^2}}\right\} + \theta nm\left(1-\mathbb{E}\left\{Q\left(\frac{\mu-R}{\delta}\right)\right\}\right)\right) \quad (40)$$

Now we need the following result.

*Proposition 2:* $\dddot{\Phi}(R) = 0$ has only one solution.
*Proof:* Obviously, $\dddot{\Phi}(0) > 0$. Letting $\dddot{\Phi}(R) = 0$ and performing a simple computation, we have

$$-\frac{\mathbb{E}\left\{\frac{1}{\sqrt{2\pi}\delta}e^{-\frac{(\mu-R)^2}{2\delta^2}}\frac{\mu-R}{\delta^2}\right\}}{\mathbb{E}\left\{\frac{1}{\sqrt{2\pi}\delta}e^{-\frac{(\mu-R)^2}{2\delta^2}}\right\}}$$
$$= \theta nm\left(2 + \theta nm\frac{1-\mathbb{E}\left\{Q\left(\frac{\mu-R}{\delta}\right)\right\}}{\mathbb{E}\left\{\frac{1}{\sqrt{2\pi}\delta}e^{-\frac{(\mu-R)^2}{2\delta^2}}\right\}}\right)\frac{e^{-\theta nmR}}{1-e^{-\theta nmR}}. \quad (41)$$

First, we can show that the left-hand side (LHS) of (41) is an increasing function in $R$. Let

$$g(R) = \mathbb{E}\left\{\frac{1}{\sqrt{2\pi}\delta}e^{-\frac{(\mu-R)^2}{2\delta^2}}\right\}. \quad (42)$$

$\frac{1}{\sqrt{2\pi}\delta}e^{-\frac{(\mu-R)^2}{2\delta^2}}$ is log-concave function [9]. Since the nonnegative weighted sum of concave functions is concave, $g(R)$ is log-concave function [9]. And hence $-\log_e g(R)$ is convex function. Note that

$$\text{LHS} = \frac{d}{dR}\left(-\log_e g(R)\right) \quad (43)$$

thus the derivative of LHS of (41) is greater than 0, and as a result it is increasing in $R$.

Next, we can prove that the right-hand side (RHS) of (41) is a decreasing function in $R$. Note that $\frac{e^{-\theta nmR}}{1-e^{-\theta nmR}}$ is decreasing in $R$. Let

$$1-\mathbb{E}\left\{Q\left(\frac{\mu-R}{\delta}\right)\right\} = \mathbb{E}\left\{\int_{-\infty}^{\frac{\mu-R}{\delta}}\frac{1}{\sqrt{2\pi}}e^{-\frac{t^2}{2}}dt\right\} = g(u(R)), \quad (44)$$

where $g(x) = \mathbb{E}\left\{\int_{-\infty}^{x}\frac{1}{\sqrt{2\pi}}e^{-\frac{t^2}{2}}dt\right\}$ and $u(x) = \frac{\mu-R}{\delta}$. We know that $g(x)$ is a log-concave function [9], and from [9, Eq. 3.10], we can see that $\log_e g$ is concave function, and $u$ is concave and nonincreasing in $R$, and hence $\log_e g(u(R))$ is concave function in $R$ directly. So $\frac{\mathbb{E}\left\{\frac{1}{\sqrt{2\pi}\delta}e^{-\frac{(\mu-R)^2}{2\delta^2}}\right\}}{1-\mathbb{E}\left\{Q\left(\frac{\mu-R}{\delta}\right)\right\}}$ is an increasing function, i.e., $\frac{1-\mathbb{E}\left\{Q\left(\frac{\mu-R}{\delta}\right)\right\}}{\mathbb{E}\left\{\frac{1}{\sqrt{2\pi}\delta}e^{-\frac{(\mu-R)^2}{2\delta^2}}\right\}}$ is a decreasing function in $R$. Thus, the RHS of (41) is decreasing in R, and hence (41) has only one solution. $\square$

Denote the unique solution to $\ddot{\Phi}(R) = 0$ as $R'$. We know that $\ddot{\Phi}(R) > 0$ for all $R < R'$, or $\dot{\Phi}(R)$ is increasing equivalently, and $\ddot{\Phi}(R) < 0$ for all $R > R'$, or $\dot{\Phi}(R)$ is decreasing equivalently. Note here that $\int_0^\infty \dot{\Phi}(R)dR = \Phi(\infty) - \Phi(0) = 0$, $\dot{\Phi}(0) = -\theta nm(1-\mathbb{E}\left\{Q(\frac{\mu}{\delta})\right\}) < 0$, so $\dot{\Phi}(R') > 0$. Otherwise, $\dot{\Phi}(R)$ is decreasing for $R > R'$, and hence $\dot{\Phi}(R) \leq 0$, $\int_0^\infty \dot{\Phi}(R)dR < 0$. Contradiction. Also note that $\dot{\Phi}(\infty) = 0$, so $\dot{\Phi}(R) > 0$ for $R > R'$. Thus, there is only solution to $\dot{\Phi}(R) = 0$ in the range $R \in (0, R')$, and $\Phi(R)$ is minimized at this value.